\documentclass[aip,imp,amsmath,amssymb,reprint,superscriptaddress]{revtex4-1}

\usepackage{graphicx}
\usepackage{dcolumn}
\usepackage{bm}
\usepackage{ulem}

\begin{document}
	
\title{Real time demonstration of high bitrate quantum random number generation with coherent laser light}

\author{T. Symul}

\author{S. M. Assad}

\author{P. K. Lam}
\email{Ping.Lam@anu.edu.au}
\affiliation{Centre for Quantum Computation and Communication Technology, Department of Quantum Science,
Australian National University, Canberra, ACT 0200, Australia}

\date{\today}

\begin{abstract}
{We present a random number generation scheme that uses broadband
  measurements of the vacuum field contained in the radio-frequency
  sidebands of a single-mode laser.  Even though the measurements may
  contain technical noise, we show that suitable algorithms can
  transform the digitized photocurrents into a string of random
  numbers that can be made arbitrarily correlated to a subset of the
  quantum fluctuations (High Quantum Correlation regime) or
  arbitrarily immune to environmental fluctuations (High Environmental
  Immunity).  We demonstrate up to 2 Gbps of real time random number
  generation that were verified using standard randomness tests.}
\end{abstract}
\maketitle

Reliable and unbiased random numbers (RNs) are needed for a range of applications spanning from numerical modeling to cryptographic communications. With the numerous improvements in quantum key distribution (QKD) protocols \cite{DVQKD, CVQKD}, fast and reliable RN generation is now one of the main technical impediment to high-speed QKD. Whilst there are algorithms that can generate pseudo-RNs, they can never be perfectly random nor indeterministic. True RNs from physical processes may offer a surefire solution.

Several physical RN generation schemes have been proposed and demonstrated \cite{Isida, Uchida, Reidler09}, including schemes based on single photon detections\cite{Rarity, Dultz, Stefanov, Jennewein, Pironio,Shields}. The limit in speed of these systems are in the dead time of photon counters.  An alternative quantum approach to photon counting is to use the vacuum fluctuations of an electromagnetic field for RN generation \cite{Trifonov, Leuchs}. In this letter, we demonstrate a simple scheme to measure and convert vacuum field fluctuations into RNs.

\begin{figure*}[!ht]
  \centering
\includegraphics[width=14cm]{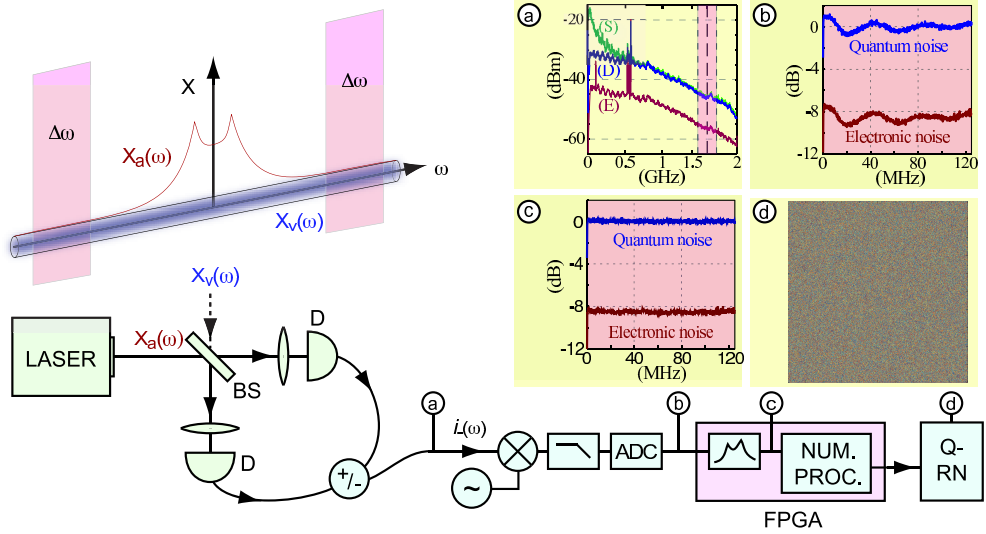}
 \caption{Random number generation schematic.  Top left figure shows the spectra
   of the quadrature amplitudes for the vacuum fluctuations, $X_v$ and
   the laser mode $X_a$.  The shaded frequency range, $\Delta \omega$,
   shows a region where the laser source is quantum noise limited.
   (a) Sum (S) and difference (D), of the laser vacuum fluctuations,
   as well as electronic noise (E) from a pair of homodyne
   photodetectors. The shaded frequency range is used for generating random numbers. (b)  Digitized and demodulated shot noise and electronic noise spectra. (c)  A filter function is used to correct for the non-uniform electronic gain. The quantum noise at this stage has an unbiased Gaussian distribution with a normalized mean at 0~dB and dark noise clearance of 8.5~dB.  (d) High randomness of the final digital random numbers depicted by the featureless color plot. BS: 50/50 beamsplitter; D: photodiode; ADC: Analog to digital converter; NUM. PROC.: Numerical processing algorithm; Q-RN: Final generated random numbers.}
  \label{fig:setup}
\end{figure*}

The schematic of the quantum RN generator is shown in Fig.~\ref{fig:setup}.  A single-mode laser beam at 1550~nm is used as the light source.  A few mW of light is split into two equal intensity beams and detected by a pair of photodetectors in a balanced homodyne scheme.  When the average laser field amplitude $\alpha$ is significantly larger than the vacuum field fluctuation the subtracted photo-current from the pair of detectors is proportional to $\alpha X_v(\omega)$,  where $X_v$ is the quadrature amplitude of the vacuum field.  The balanced homodyne setup therefore measures the amplified quadrature amplitude of the vacuum field fluctuations.  Only sideband frequencies well above the technical noise frequencies of the laser are used for RN generation (shaded region of the radio frequency (RF) spectrum of Fig.~\ref{fig:setup}(a)). This is achieved by demodulating the photocurrent with an RF frequency (1.6 GHz) followed by a low pass filter.  The undulations in the spectra are due to non-uniform RF electronic gain in the photodetectors amplification stages (Fig.~\ref{fig:setup}(b)).  Nevertheless, the quantum noise has a constant clearance above the electronic noise level of 8.5 dB.  Using a Field-programmable Gate Array (FPGA) a filter function can be programmed to neutralize the non-uniform electronic gain as shown in  Fig.~\ref{fig:setup}(c). Finally, using suitable numerical processes, the quantum noise is converted into a sequence of random digital bits as depicted by the 8-bits colour code in Fig.~\ref{fig:setup}(d).

In practice, vacuum fluctuations cannot be detected in complete
isolation. The electronic noise of the photo-detector will be
superimposed onto the measured photocurrents.  While it is reasonable
to assume that the quantum noise of a vacuum field $X_v$ is perfectly random over all frequencies and cannot be tampered with, electronic noise may not possess these ideal properties. For this letter, we would like to consider two possible scenarios.  

In the first scenario, we assume that the electronic noise is untampered.  We wish to find a protocol to generate RNs solely from closely tracking the quantum fluctuations of the vacuum field.  We will show that in this scenario, a 1-bit digitized encoding of the vacuum fluctuations, together with thresholding, can indeed allow arbitrarily {\it high quantum correlation}. In the second scenario we assume that the electronic noise may be tampered and is untrustworthy.  We desire an algorithm that will generate RNs that are tamper-proof, even in the presence of possibly forged electronic noise.  Provided that quantum noise remains the dominant source of noise, we will show that digitizing the vacuum fluctuations into multiple bits, and then discarding the most-significant bits, can indeed allow arbitrarily {\it high environmental immunity}. Correlations between environmental noise and the generated RNs can be made arbitrarily small.

\begin{figure}[!ht]
  \centering
\includegraphics[width=\columnwidth]{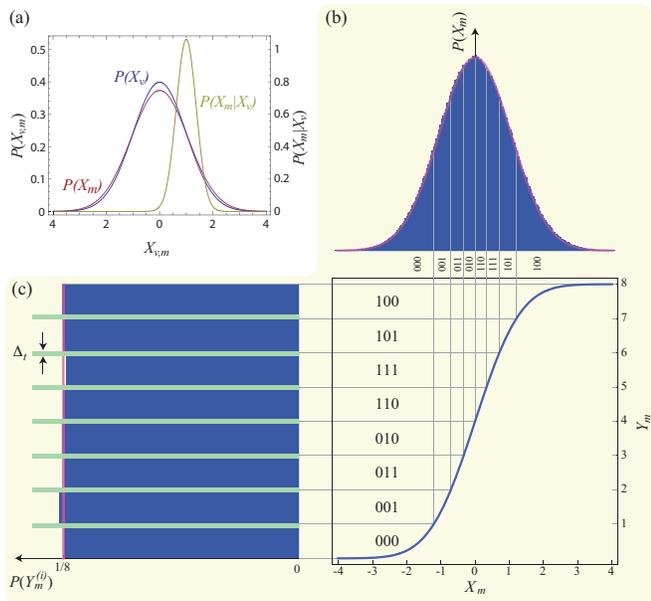}
  \caption{(a) Gaussian distribution of the measured and vacuum fluctuations $P(X_{m,v})$, and the conditional probability distribution of  $P(X_m | X_v)$ for an arbitrary value of $ X_v=1$.  The figure is plotted for a $X_e$ that is 8.5~dB below $X_v$.  (b) 1G samples of the Gaussian distributed $X_m$ after 12-bits digital filtering.  Experimental data is plotted with blue bars and its corresponding theoretical expectation with purple lines. (c) The Gaussian error function is used to transform $X_m$ into a uniform distribution of $P(Y_m)$. Thresholding (represented by the green bars) can be introduced to reject data points that fall within a $\Delta_t$ range from the boundary of two adjacent domains.}
  \label{fig:uni}
\end{figure}

We describe the measured signal as $X_m = X_v + X_e$, where $X_e$ corresponds to the electronic noise superimposed onto the vacuum fluctuations $X_v$ (see Fig. \ref{fig:uni}(a)).  $X_e$ and $X_v$ can be modeled as two uncorrelated Gaussian distributions of zero mean and variance $V_e$ and $V_v$, respectively.
$X_m$ is also Gaussian with a conditional probability given by
\begin{equation}
P(X_m | X_v)=\frac{1}{\sqrt{2 \pi V_e}} e^{-\frac{(X_m-X_v)^2}{2 V_e}}\;.
\end{equation}
In order to extract $n$-bit RNs from $X_m$ with uniform probabilities, we need to transform the measured Gaussian distributed photocurrent into a uniform distribution of $Y_m= \left[1+{\rm erf}(X_m/\sqrt{2 V_m})\right]/2$. The $n$-bit RNs require the distribution $Y_m$ to be divided into $2^n$ equal and non-overlapping domains (as shown in Fig.~\ref{fig:uni} for the case of $n=3$).  We can then index each domain, preferably by using Gray's binary encoding \cite{Gray}, to minimize the perturbations due to unaccounted technical noise so that the $2^n$ domains now correspond directly to the $n$-bits RNs (see Fig.~\ref{fig:uni}(b)).

We introduce a {\it thresholding} condition to reject data points that fall within a certain range $\Delta_t$ between two adjacent domains (see Fig.~\ref{fig:uni}). We digitize continuous variable quantities into a binary $n$-bits with $Y_m = (Y_m^{(1)}Y_m^{(2)}...Y_m^{(n)}) $, where $Y_m^{(i)} \in \{ 0,1 \}$.  $Y_m^{(1)}$ is the most significant bit (MSB) and $Y_m^{(n)}$ the least significant bit (LSB) in Gray's binary decomposition.
We define a probability of error, $P_{e,q}  = P( Y_m^{(i)} \neq Y_v^{(i)}) $. This quantifies the discrepancy between a measured bit and the vacuum fluctuations.  Perturbation from the electronic noise, which causes the digitized bit to differ from the vacuum fluctuation, is referred to as an error.
$P_{e,q} = 0$ means that the generated RNs are the perfect digitization of pure vacuum fluctuations.

We also introduce the notion of information leakage from the electronic noise $I_{\rm E}$.  This information leakage is the amount of information that can be imposed on the final random bits by somebody having access to the electronic noise.  Using Shannon's binary information formula, we obtained
\begin{equation}
I_{\rm E}=1+P_{e,e} \log_2\left(P_{e,e}\right)+\left(1-P_{e,e}\right) \log_2\left(1-P_{e,e}\right)
\end{equation}
where analogous to $P_{e,q}$, $P_{e,e}$ is the probability of error that the measured $i^{\rm th}$-bit differs from the electronic noise. $I_{\rm E} = 0$ means that the generated RNs are indeterminable even with full knowledge of the electronic noise.

\begin{figure}[!ht]
\includegraphics[width=\columnwidth]{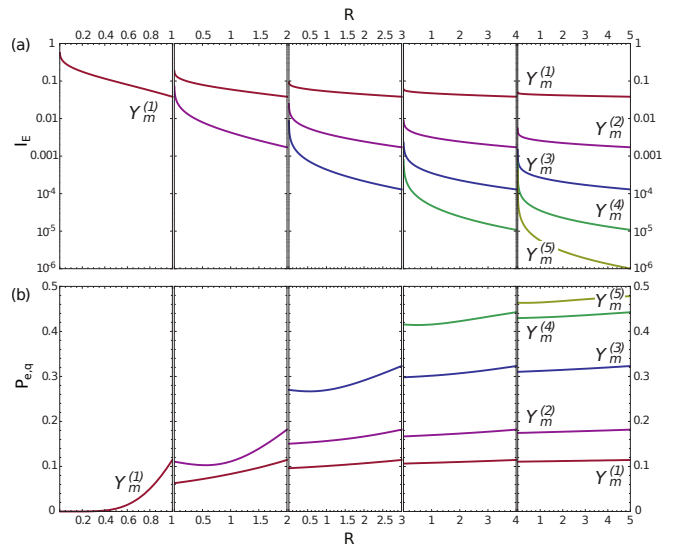}
  \caption{(a) Information leakage, $I_{\rm E}$, for multi-bits division of vacuum fluctuations as a function of RN generation rate, $R$ corresponding to the number of bits extracted times the probability of not being rejected by the thresholding procedure.  The LSB from the Gray code shows the least amount of information leakage.  (b)  Probability of error, $P_{e,q}$, plotted as a function of $R$.  Fewer bits encoding gives smaller $P_{e,q}$.}
\label{fig:MSBLSB}
\end{figure}

Figure~\ref{fig:MSBLSB} shows $I_{\rm E}$ and $P_{e,q}$ plotted as a function of the RN generation rate $R = n[1-P_{\rm thr}(\Delta_t)]$ for $n=1,2,3,4,5$ bits extracted per measurement, where $P_{\rm thr}(\Delta_t)$ is the probability that a bit is rejected because it lies within one of the thresholding areas. From Fig.~\ref{fig:MSBLSB}(a) we note that the amount of information leaked through the electronic noise decreases for the least significant bit (LSB) with high bit encoding.  Moreover, thresholding paradoxically increases $I_{\rm E}$.  Minimizing $I_{\rm E}$ therefore requires high bit encoding, no thresholding, and omission of all but the LSB.  In the case of 5-bits encoding, the best result obtained using this protocol was $I_{\rm E} < 10^{-6}$ for the LSB.

In Fig.~\ref{fig:MSBLSB}(b) we see that the MSBs are the most accurate
representation of the vacuum fluctuations.  In contrast to information
leakage, thresholding in general reduces the probability of error
$P_{e,q}$ since thresholding effectively discards all data where
electronic noise is the dominant source of error, leaving behind only data with large vacuum fluctuations.  Thresholding is therefore favorable for increasing the correlations between the RNs and the vacuum fluctuations.  For multi-bit rate encoding, however, $P_{e,q}$ is nonzero even with very large thresholding.  In fact, at high bit rates $P_{e,q}$ is no longer critically dependent on thresholding.  An ideal protocol for maximizing the quantum correlation of RN is therefore a single bit encoding with threshold value significantly larger than the electronic noise. From this we can define two regimes of operation for our quantum RN generator: (i) A regime of {\it high quantum correlation} where vacuum fluctuations are accurately converted to digital RNs. (ii) A regime of {\it high environmental immunity} where tampering of photodetector electronic noise does not compromise the indeterminacy of the RN generation.

We implemented our proposed algorithm in real-time using an integrated
12 bits 250 Msamples per second analog to digital converter (ADC) and a FPGA.
Our results show a uniformly distributed random binary sequences where 8 bits are extracted for each measurement without thresholding, corresponding to a real-time random bit rate generation of 2~Gbps.  This demonstration consistently passes the NIST \cite{NIST} and Diehard randomness tests \cite{DH}.

In the context of the two previously introduced regimes, we also implemented a {\it `high quantum correlation'} single-bit encoding with thresholding rejecting 90\% of the samples, and a {\it `high environmental immunity'} zero threshold 8-bit encoding that only keeps the 4 LSBs.  In the {\it `high quantum correlation'} mode of operation the binary random sequence is produced at a rate of 25~Mbps, with $P_{e,q}<10^{-6}$, whilst in the {\it `high environmental immunity'} regime the random bit-rate is 1~Gbps with $I_E<10^{-6}$.

In conclusion, we have demonstrated the generation of continuous random bit sequences at a rate of 2~Gbps in real time  by sampling the broadband vacuum fluctuations.  We proposed two methods of generating RNs where: (i) {\it quantum correlation} is optimized for a near ideal representation of a thresholded random subset of the vacuum fluctuations, and (ii) {\it environmental immunity} is optimized to combat against possible tampering of the electronic noise. 

We thank QuintessenceLabs, M. Neharkar and K. L. Chong for technical assistance.  This research was conducted by the Australian Research Council Centre of Excellence for Quantum Computation and Communication Technology (project number CE110001029).

\end{document}